\documentclass[lettersize,journal]{IEEEtran}
\usepackage{amsmath,amsfonts}
\usepackage{algorithmic}
\usepackage{algorithm}
\usepackage{array}
\usepackage[caption=false,font=normalsize,labelfont=sf,textfont=sf]{subfig}
\usepackage{textcomp}
\usepackage{stfloats}
\usepackage{url}
\usepackage{verbatim}
\usepackage{graphicx}
\usepackage{cite}
\usepackage{xcolor,cite,etoolbox}
\begin{document}
\title{Plasma-Enabled Wideband Matching of Electrically Small Antennas}

\author{Kushagra Singhal, \IEEEmembership{Graduate Student Member, IEEE}, and Abbas Semnani, \IEEEmembership{Senior Member, IEEE}

\thanks{The authors are with the Department of Electrical Engineering and Computer Science, The University of Toledo, Toledo, Ohio 43606, USA. (e-mail: kushagra.singhal@rockets.utoleod.edu; abbas.semnani@utoleod.edu). This work was supported by the Office of Naval Research (ONR) under Grant N00014-21-1-2449.}}

\markboth{IEEE Transactions on Antennas and Propagation,~Vol.~x, No.~xx, 2026}%
{Shell \MakeLowercase{\textit{et al.}}: A Sample Article Using IEEEtran.cls for IEEE Journals}

\maketitle

\begin{abstract}
This paper presents a plasma-based approach for wideband impedance matching of electrically small antennas (ESAs). By exploiting the negative permittivity of a capacitively coupled plasma (CCP), the plasma is shown to function as a wideband negative capacitor that effectively compensates for the large and frequency-dependent reactance inherent to ESAs. A VHF/UHF electrically small capped dipole antenna is used as a proof-of-concept platform to demonstrate the proposed technique. Theoretical analysis, Multiphysics simulations, and experimental measurements confirm broadband impedance matching, with -6 dB fractional bandwidth that can exceed 100\%, using only a few watts of plasma power. The matching bandwidth can be dynamically tuned by adjusting the plasma operating conditions, gas pressure, and electron density, enabling control over the trade-off between bandwidth and radiation efficiency. This plasma-enabled matching paradigm provides a physically robust and scalable solution for wideband ESAs, particularly attractive for low-frequency and high-power applications where conventional matching techniques are fundamentally limited.
\end{abstract}


\section{Introduction}  
\label{sec:introduction}
\IEEEPARstart{E}{lectrically} small antennas (ESAs) have long attracted significant interest within the scientific and engineering communities. The ongoing trend toward miniaturization and multifunctionality in portable and wireless systems, coupled with increasing operational frequency requirements, has intensified demand for compact antennas capable of broadband operation. In addition, many practical applications, particularly at low radio frequencies, impose severe constraints on the available antenna volume, further motivating the development of wideband ESAs. An antenna is classified as electrically small when its electrical size, defined by $ka$, is less than unity, where $k$ is the free-space wavenumber and $a$ is the radius of the smallest sphere that surrounds the antenna. Due to their compactness, conventional ESAs are fundamentally limited in bandwidth and radiation efficiency and are notoriously difficult to match over wide frequency ranges \cite{TAP1_1,TAP1_2}. As a result, practical implementations often rely on suboptimal designs that require severe performance trade-offs.

The input impedance of an ESA is typically characterized by a small radiation resistance accompanied by a large, strongly frequency-dependent reactance, underscoring the central role of impedance matching \cite{TAP1_2}. However, achieving wideband matching with conventional passive networks is fundamentally constrained by the gain–bandwidth limitation \cite{TAP1_1}. Consequently, substantial effort has been devoted to identifying non-LTI (linear time-invariant) strategies capable of providing broadband impedance matching, ideally canceling the antenna reactance while preserving an appropriate resistive component. One promising class of approaches exploits non-Foster matching networks that emulate negative reactance, most commonly negative capacitance.

Negative impedance converters (NICs) have been extensively investigated for this purpose \cite{TAP1_5,TAP1_6,TAP1_9,TAP1_10}. In parallel, metamaterial-inspired ESAs have emerged for antenna miniaturization and bandwidth enhancement. These approaches can be broadly grouped into four categories \cite{lit6_1}. The first comprises dispersion-engineered antennas, most notably composite right/left-handed (CRLH) transmission-line structures \cite{lit6,lit7,lit10}, which support zeroth- and negative-order resonant modes through tailored dispersion characteristics. Because their resonance frequencies are weakly dependent on physical dimensions, CRLH-based antennas enable substantial miniaturization without severe degradation of radiation performance.

The second category includes metamaterial-loaded antennas that employ epsilon-negative (ENG) and mu-negative (MNG) media \cite{TAP1_11,TAP1_12,lit12,lit13}, as well as high-permeability shells and magnetic photonic crystals \cite{lit14,lit16}, which effectively reduce the electrical size of the antenna by confining electromagnetic energy. Closely related are meta-resonator-based antennas that utilize subwavelength inclusions, such as split-ring resonators (SRRs) and complementary SRRs (CSRRs) \cite{lit17,lit18,lit20}, which offer deep miniaturization, tunability, and compatibility with planar fabrication processes. The fourth category comprises metasurface-loaded antennas, including electromagnetic bandgap (EBG) and reactive impedance surfaces (RIS) \cite{lit22,lit23,lit24}, which impose engineered boundary conditions to suppress surface waves and enhance radiation efficiency.

More recently, modulation-based techniques have been proposed to address the limitations of LTI matching networks by dynamically manipulating the antenna's reactive boundary conditions. Direct antenna modulation (DAM) \cite{lit28} enables broadband operation by rapidly switching antenna load in synchrony with the RF waveform, thereby preserving near-field energy and minimizing switching losses. However, DAM implementations are constrained by transistor parasitics, stringent timing requirements, and efficiency degradation associated with finite switch ON-resistance.

An alternative modulation-based approach employs space–time modulation to induce controlled multimode coupling in sectorized ESA architectures \cite{lit29}. In particular, a four-sector top-hat-loaded monopole supports one radiating mode and several high-$Q$ non-radiating (dark) modes. By applying spatially discrete traveling-wave modulation (SDTWM) using time-varying capacitors with prescribed phase progression, parametric energy exchange is enabled among these modes. When properly designed, this interaction significantly broadens the effective matching bandwidth while inherently suppressing spurious radiation by channeling unwanted harmonics into non-radiating modes, eliminating the need for external idler resonators. Although DAM and SDTWM demonstrated the potential of time-varying and spatio-temporal modulation for broadband ESA matching, these systems often suffer from increased complexity, loss, stability concerns, and reduced efficiency.

Plasma offers a fundamentally distinct and physically compelling alternative, as it is the only naturally occurring medium capable of exhibiting negative permittivity. This unique property enables plasma-based matching networks to directly compensate for the large capacitive reactance inherent to ESAs. Plasma-assisted impedance matching has previously been demonstrated; however, prior implementations relied on weakly collisional plasmas of very low pressure (3–10 mTorr) that behaved predominantly as tunable inductive elements \cite{plasESA1,plasESA2}. In such cases, matching was achieved through inductive resonant compensation of the antenna's capacitive reactance, resulting in narrowband, albeit tunable, operation.

In this work, we investigate mid-collisional capacitively coupled plasmas as broadband negative-capacitance elements for ESA matching. In this case, an RF electric field efficiently accelerates electrons within the plasma sheath, enabling substantial energy exchange while avoiding the strict resonance conditions of collisionless operation. Although finite electron–neutral momentum-transfer collisions introduce ohmic losses, the resulting plasma response enables wideband reactance compensation with large fractional bandwidth (FBW). This approach relaxes the fundamental bandwidth constraints associated with resonant matching and provides an efficient, physically robust pathway to wideband impedance matching and enhanced radiation performance in ESAs.

\begin{figure*}[!]
    %\centering
    \includegraphics[width=0.98\textwidth]{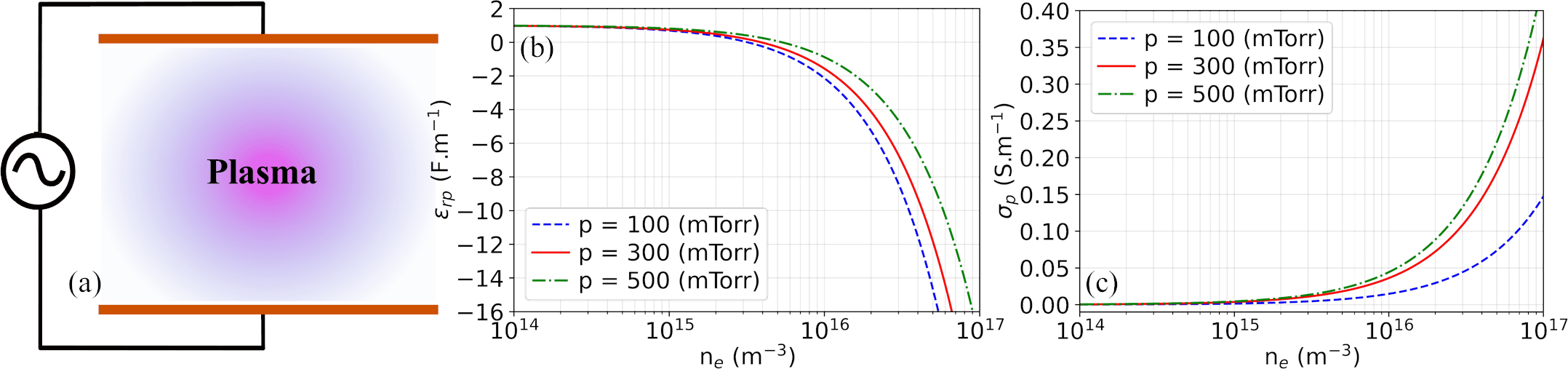}
    \caption{(a) Capacitively coupled plasma (CCP) configuration. (b) Variation of plasma relative permittivity and (c) conductivity as functions of electron number density at 500 MHz for different argon pressures, illustrating the transition to negative permittivities at moderately low plasma densities.}
    \label{fig:1}
\end{figure*}

\section{Theory of Plasma Matching}

\label{sec:Theory}
In the absence of an external magnetic field, a weakly ionized plasma can be accurately modeled using the Drude approximation as a dispersive, lossy dielectric medium characterized by complex permittivity and finite conductivity \cite{TAP1_13,TAP1_15}. In this model, plasma relative permittivity \((\epsilon_{rp})\) and conductivity are given by \((\sigma_p)\)
\begin{equation}
\epsilon_{rp} = 1 - \frac{e^2 n_e}{\epsilon_0 m_e \left(\omega^2 + \nu_m^2 \right)},
\end{equation}
\begin{equation}
\sigma_p = \frac{e^2 n_e \nu_m}{m_e \left(\omega^2 + \nu_m^2 \right)},
\end{equation}
where \(\epsilon_0\) is the free-space permittivity, \(m_e\) is the electron mass, \(n_e\) is the electron number density, \(\omega\) is the angular frequency, and \(\nu_m\) is the electron–neutral momentum-transfer collision frequency, which can be estimated as
\begin{equation}
\nu_m \approx C_{\text{gas}} \times p,
\end{equation}
where \(p\) is the gas pressure and \(C_{gas}\) is a gas-dependent constant, listed in Table.~\ref{tab1}.

Figures~\ref{fig:1}(b) and (c) illustrate the dependence of plasma relative permittivity and conductivity on the electron number density for argon at different pressures. These results highlight the dual dielectric–conductive nature of plasma. Notably, the relative permittivity remains below unity and can become negative over a broad range of plasma densities, a property that is central to plasma-enabled impedance matching. Among plasma generation techniques, capacitively coupled plasma (CCP) is particularly attractive owing to its simplicity and compatibility with RF systems. In a CCP configuration, an externally applied RF electric field ionizes the neutral gas within a capacitive structure, producing a plasma discharge \cite{PlasmaDischarge}. By carefully controlling the plasma excitation conditions to achieve a desired electron density while limiting conductive losses, the effective plasma permittivity can be tailored to realize a negative capacitive response, which is critical for compensating the strongly capacitive impedance of ESAs.

\begin{figure}[!]
    \centering
    \includegraphics[width=0.8\columnwidth]{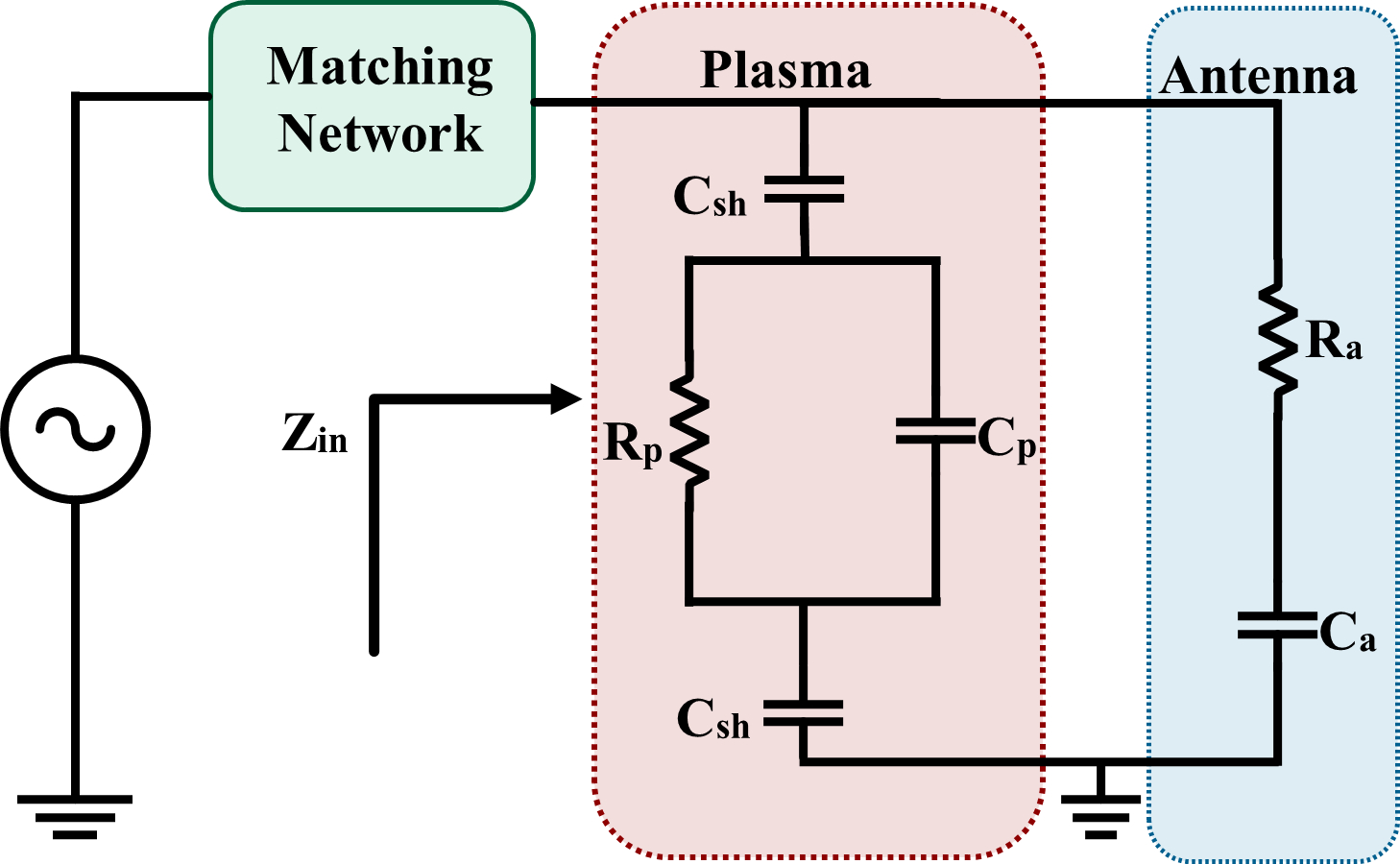}
    \caption{Equivalent circuit diagram of the proposed plasma-matching technique for electrically small antennas, in which a negative-permittivity CCP is connected in parallel with the ESA to compensate its reactance over a wide bandwidth.}
    \label{fig:2}
\end{figure}

The proposed plasma-based wideband impedance-matching concept is illustrated in Fig.~\ref{fig:2}, where a CCP element is integrated in parallel with an ESA. To analyze the electrical behavior of the plasma discharge, a lumped-element equivalent circuit is adopted. In this model, the bulk plasma is represented by a parallel $RC$ network, whereas two series capacitances model the plasma sheaths adjacent to the electrodes. 
Plasma resistance $R_p$ and capacitance $C_p$ are defined as
\begin{equation}
R_p = \frac{d}{\sigma_p A_p}, \qquad
C_p = \frac{\epsilon_0 \epsilon_{rp} A_p}{d},
\end{equation}
where $d$ is the characteristic discharge gap length and $A_p$ is the effective cross-sectional area of the plasma volume. 
The admittance of the bulk plasma can be expressed as
\begin{equation}
\frac{1}{Z_p} = \frac{1}{R_p} + j\omega C_p,
\end{equation}
and the total CCP impedance is given by
\begin{equation}
Z_{\text{ccp}} = Z_p - \frac{2j}{\omega C_{\text{sh}}}.
\end{equation}

The objective of the plasma-based matching concept is to introduce a capacitive plasma region whose net reactance cancels the ESA's capacitive reactance over a wide bandwidth. The CCP impedance is connected in parallel with the antenna impedance $Z_a$, resulting in the total input impedance
\begin{equation}
\frac{1}{Z_{\text{in}}} = \frac{1}{Z_{\text{ccp}}} + \frac{1}{Z_a},
\end{equation}
where $Z_a$ is the ESA's input impedance, modeled as
\begin{equation}
Z_a = R_a - \frac{j}{\omega C_a}.
\end{equation}
Here, $R_a$ is the resistance to radiation and $C_a$ represents the inherent capacitive reactance of the antenna. When the plasma capacitive reactance is designed as proposed, the imaginary part of $Z_{in}$ is minimized, leaving a predominantly resistive input impedance over a broad frequency range. To achieve this, the capacitive plasma section must be designed in a way that
\begin{equation}
\operatorname{Im}[\frac{1}{Z_{\text{ccp}}}] = - \operatorname{Im}[\frac{1}{Z_{\text{a}}}].
\end{equation}

To theoretically evaluate the proposed technique, a typical electrically small antenna with resistance and capacitance of $R_a = 1.2~\Omega$ and $C_a = 5.3$ pF at 400~MHz is assumed. The plasma and gas parameters used in the model are summarized in Table~\ref{tab1}. By incorporating antenna and plasma characteristics, this framework enables a quantitative evaluation of plasma-matched ESA performance under realistic operating conditions.

\begin{table}[!]
\centering
\caption{\label{tab1}Design Parameters: Circuit Model and Geometric Dimensions of the Plasma-Matched Capped Dipole ESA}
\begin{small}
\renewcommand{\arraystretch}{1.15}
\begin{tabular}{c c c c}
\hline  \hline
Parameter & Symbol  & Value & Unit \\ 
\hline
Antenna resistance   & $R_a$    & 1.2  & $\Omega$ \\ 
Antenna capacitance   & $C_a$   & 5.3 & $pF$ \\ 
Sheath thickness & $d_{sh}$  & 900 & $\mu m$ \\ 
Argon gas constant  & $C_{gas}$  & $5.3\times10^{9}$ & $s^{-1}~Torr^{-1}$ \\ 
Antenna radius  & $r_a$    & 40  & $mm$ \\ 
Plasma radius  & $r_p$    & 20 & $mm$ \\ 
Plasma gap & $d$ & 25 & $mm$ \\ 
Gas pressure  & $p$    & 300 & $mTorr$ \\ 
Glass thickness  & $t_g$  & 2 & $mm$ \\ 
\hline  \hline
\end{tabular}
\end{small}
%\end{ruledtabular}
\end{table}

\begin{figure}[!]
    \centering
    \includegraphics[width=0.95\columnwidth]{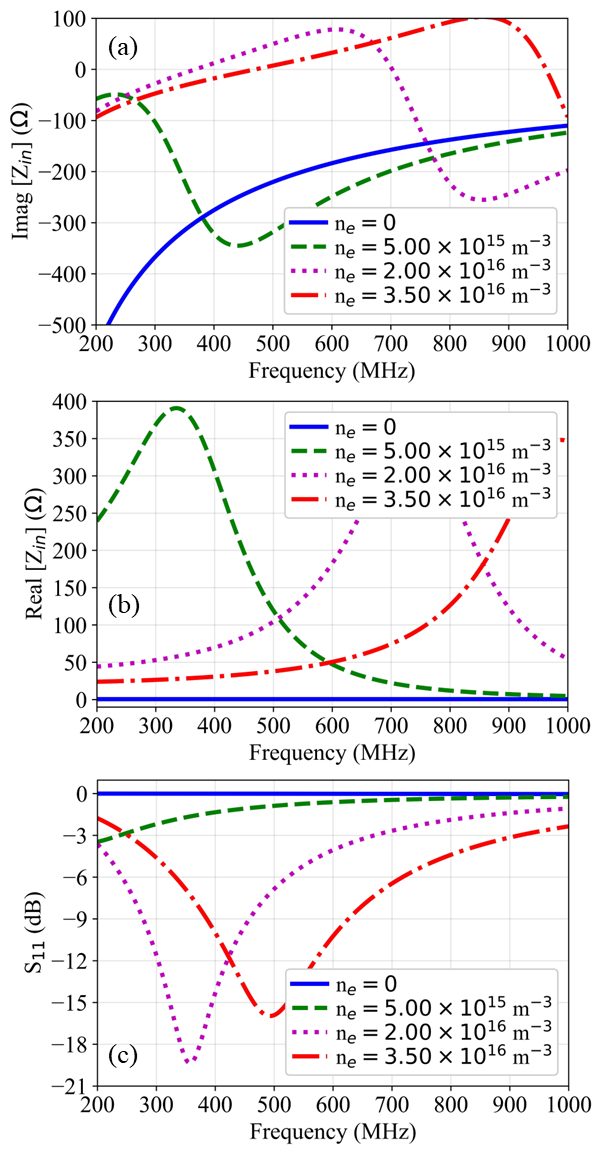}
    \caption{Theoretical evaluation of the impact of a parallel plasma with varying electron number densities on the input impedance and reflection coefficient of an electrically small antenna, demonstrating plasma-enabled wideband impedance matching.}
    \label{fig:4}
\end{figure}

In this configuration, the plasma effectively behaves as a negative capacitor, thereby counteracting the ESA's inherent capacitive reactance. Figure~\ref{fig:4}(a) demonstrates that increasing the electron number density shifts the overall reactance to zero, enabling improved impedance matching. Similarly, Fig.~\ref{fig:4}(b) shows an increase in the real part of the input impedance with the electron density, reflecting the resistive contribution of the plasma. This tunability provides direct control over both reactive and resistive impedance components, enabling a self-matching mechanism that reduces dependence on external matching networks and enhances system robustness. In addition, plasma-induced impedance shaping significantly broadens the antenna's operational bandwidth. The theoretical reflection coefficient shown in Fig.~\ref{fig:4}(c) confirms that the proposed approach maintains a reflection coefficient below -6 dB over a wide frequency range, demonstrating effective wideband matching and improved ESA performance.

\section{Design and Multiphysics Modeling}
\label{sec:Cap_Dipole}

\begin{figure}[!]
    \centering
    \includegraphics[width=0.56\columnwidth]{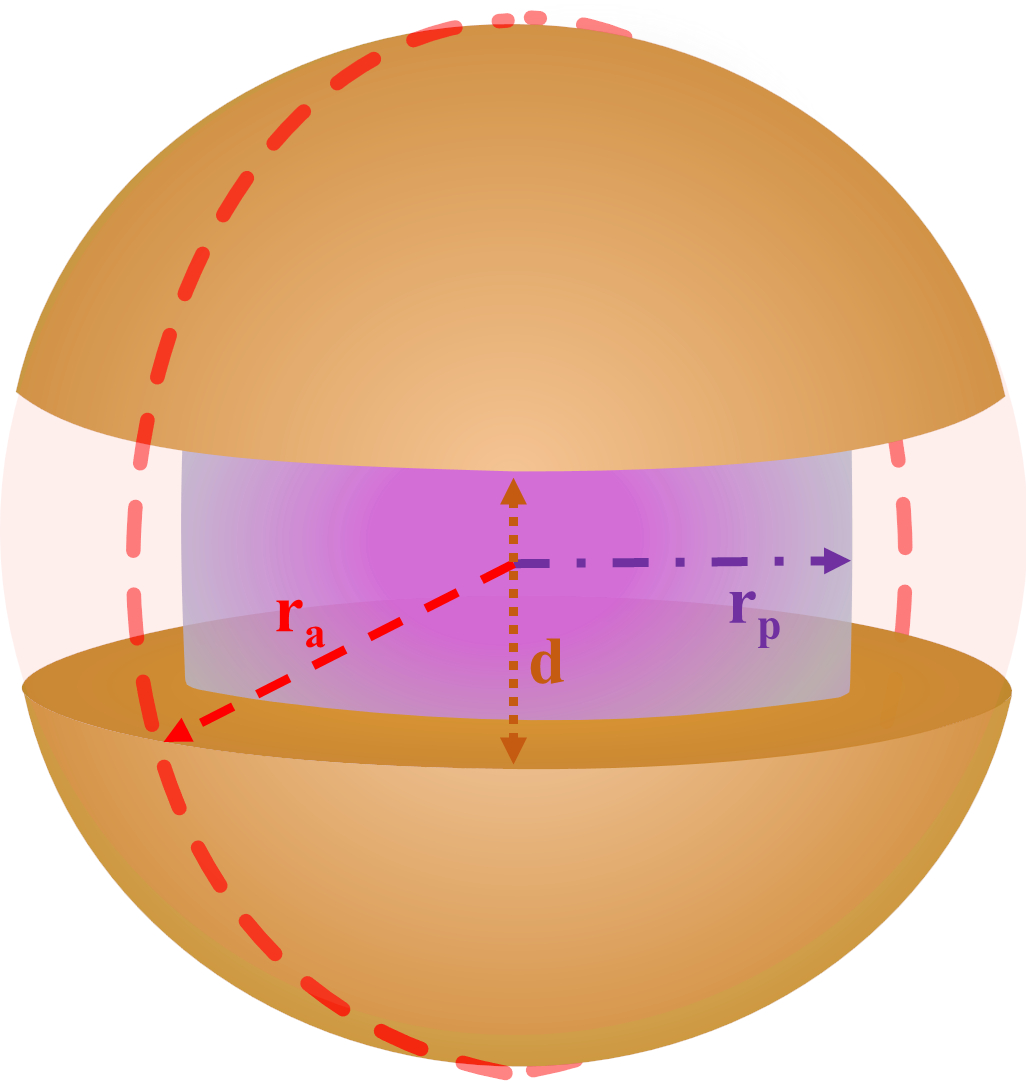}
    \caption{Proposed capped dipole ESA configuration, which efficiently utilizes the spherical volume and provides a suitable geometry for sustaining the required capacitively coupled plasma.}
    \label{fig:3}
\end{figure}

To implement the proposed plasma-matching technique, an electrically small spherical-capped dipole antenna is considered, consisting of two identical spherical domes separated by a central gap, as illustrated in Fig.~\ref{fig:3}. This geometry is selected for two primary reasons. First, maximizing radiation efficiency in an ESA requires effective utilization of the available circumscription volume \cite{TAP1_11}. Volumetric antenna configurations address this requirement by occupying a large fraction of the enclosed sphere, thus mitigating fundamental efficiency and bandwidth limitations \cite{TAP1_2}. Second, the central gap between the two spherical caps provides a large, well-defined capacitive region with a relatively uniform electric field, which is essential for reliable ignition and stable operation of a CCP. This configuration is particularly advantageous for maintaining plasma at moderate pressures with controllable electron density. The proposed antenna has an overall diameter of 80 mm, corresponding to an electrical size of $ka$ = 0.5 at 600 MHz. The detailed geometrical and operational parameters of the antenna–plasma system are summarized in Table~\ref{tab1}.

Given the inherently interdisciplinary nature of plasma-enabled antennas, a comprehensive multiphysics simulation framework was developed using COMSOL Multiphysics~6.4 \cite{comsol}. The model couples the RF Module with the Plasma Module to capture the self-consistent interaction between electromagnetic fields and plasma dynamics. A detailed argon plasma chemistry model is presented \cite{PlasmaChem3,PlasmaChem6}, which includes elastic collisions, excitation, and ionization processes. These mechanisms are essential for accurately predicting plasma generation, energy transfer, and loss mechanisms. In the coupled formulation, the electromagnetic fields are solved in the frequency domain, while plasma transport and kinetics are resolved in the time domain. The multiphysics coupling links the local RF electric fields to electron heating and ionization, while the resulting plasma permittivity and conductivity are fed back into the electromagnetic solver. This approach ensures a fully self-consistent solution that captures the bidirectional coupling between the plasma and antenna.

The antenna geometry used in the COMSOL simulations follows the configuration shown in Fig.~\ref{fig:3}. To represent a practical implementation, a dielectric glass tube is included to confine the plasma to the desired section of the central gap. The model consists of two identical spherical caps of radius $r_a$, separated by a gap $d$ partially filled with argon at 300 mTorr. The plasma is confined within a glass tube of radius $r_p$ and thickness of 2 mm. The complete set of design parameters used in the simulations is listed in Table~\ref{tab1}.

\begin{figure*}[!]
    \centering
    \includegraphics[width=0.98\textwidth]{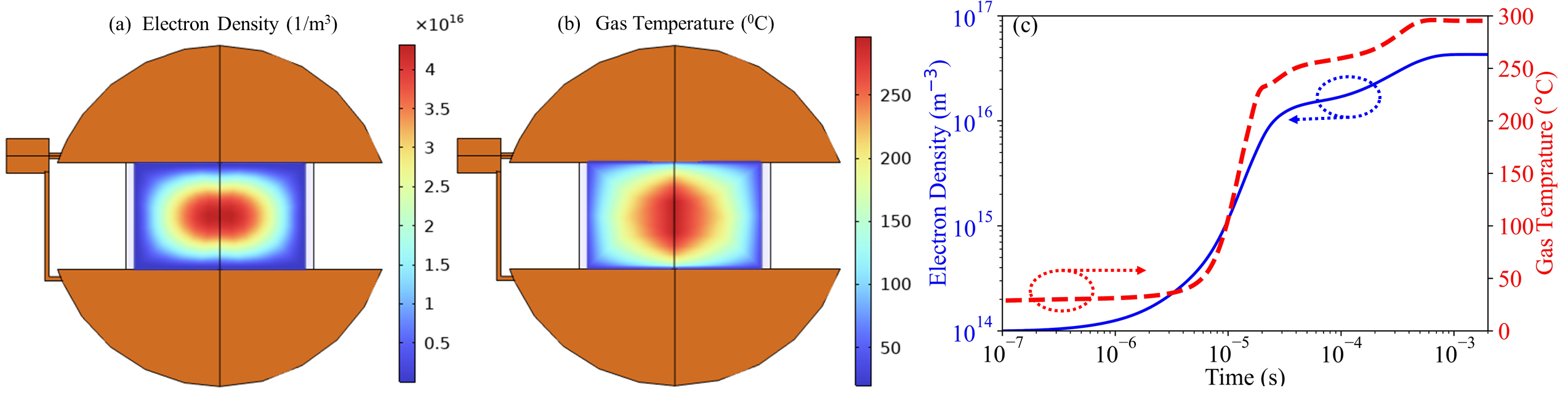}
    \caption{COMSOL Multiphysics simulation results for argon plasma at 300 mTorr with a 700 mW, 150 MHz excitation signal: (a) electron number density, (b) gas temperature, and (c) temporal evolution of electron density and temperature at the center of the plasma chamber, showing steady-state behavior on the order of hundreds of microseconds.}
    \label{fig:6}
\end{figure*}

Figure~\ref{fig:6} presents steady-state plasma simulation results for an input power of 700~mW at 150 MHz, as the RF plasma excitation frequency. Figure~\ref{fig:6}(a) shows the spatial distribution of the electron number density, while Fig.~\ref{fig:6}(b) depicts the gas temperature distribution. The maximum gas temperature inside the plasma chamber reaches approximately $300^{\circ}$ C, while the antenna structure itself remains near room temperature, indicating effective thermal isolation of the plasma region. Figure~\ref{fig:6}(c) shows the temporal evolution of electron density and gas temperature at the center of the plasma chamber, in the range of $10^{16}~m^{-3}$, demonstrating that steady state is achieved within approximately 0.5~ms.

\begin{figure}[!]
        \centering
    \includegraphics[width=0.95\columnwidth]{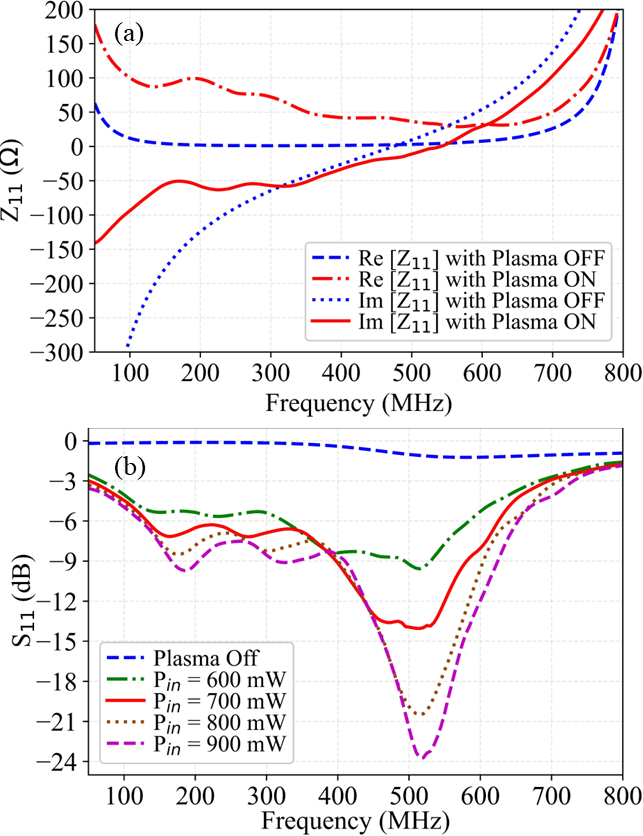}   
    \caption{COMSOL Multiphysics–simulated real and imaginary components of the antenna input impedance in the plasma-off and plasma-on operating modes.}
    \label{fig:7}
\end{figure}

The impact of plasma excitation on the antenna input impedance is illustrated in Fig.~\ref{fig:7}(a), which compares the real and imaginary components of the impedance in the plasma-off and plasma-on states. Without plasma, the antenna exhibits the characteristic behavior of an ESA, namely, low radiation resistance and strongly frequency-dependent capacitive reactance. When the plasma is ignited, its effective negative capacitance compensates for the antenna’s inherent capacitive reactance. As a result, the plasma-augmented antenna exhibits a near-zero reactive component and a resistive input impedance approaching 50~$\Omega$ over a broad frequency range.

Figure~\ref{fig:7}(b) shows the simulated reflection coefficient of the antenna with and without plasma for different levels of plasma excitation power. In the plasma-on state, the reflection coefficient remains below -6 dB over the frequency range from 200 to 700 MHz, indicating a substantial improvement in impedance matching. Moreover, the results demonstrate that the matching bandwidth can be tuned by adjusting the plasma input power. Higher plasma densities yield broader impedance matching at the expense of increased conductive losses, thereby reducing radiation efficiency.

\begin{figure}[!]
    \centering
     \includegraphics[width=0.95\columnwidth]{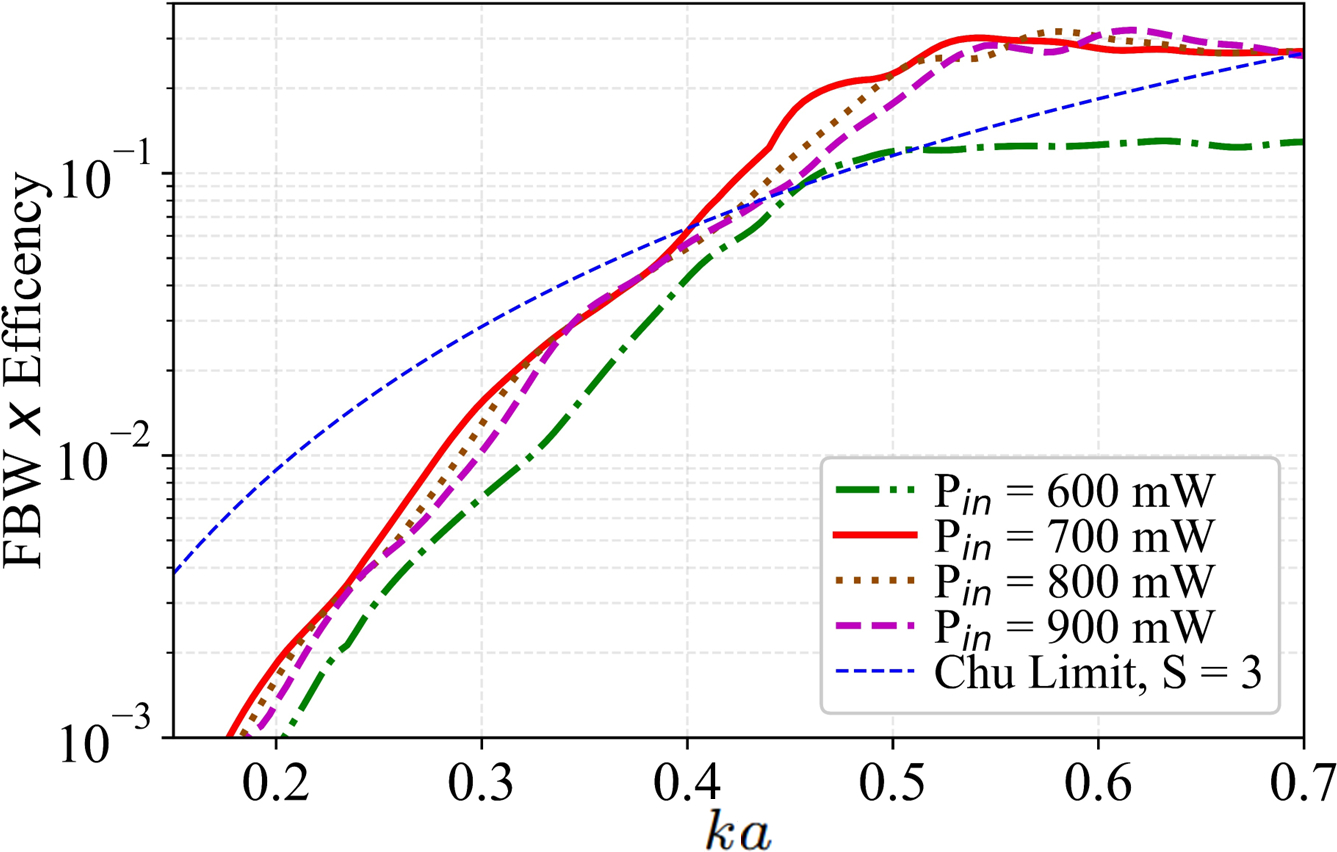}    
    \caption{COMSOL-simulated product of bandwidth and radiation efficiency as a function of $ka$ across the electrically small antenna regime, demonstrating performance exceeding the Chu limit for $ka > 0.4$.}
    \label{fig:9}
\end{figure}

\begin{figure}[!]
    \centering
    \includegraphics[width=0.98\columnwidth]{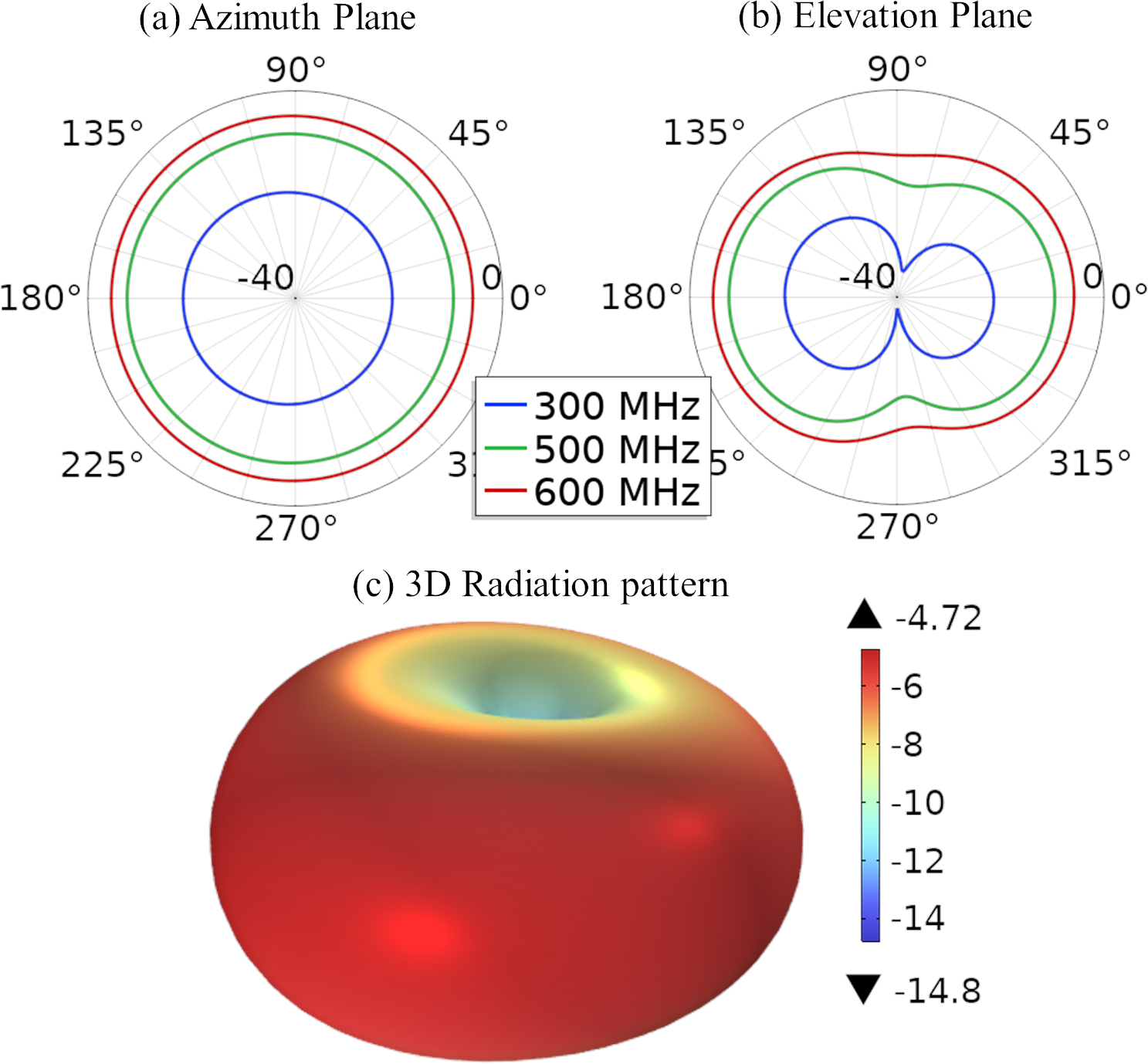}
    \caption{COMSOL-simulated radiation patterns and realized gain at three frequencies in (a) the azimuth plane and (b) the elevation plane, along with (c) the three-dimensional radiation pattern at 600 MHz ($ka$ = 0.5), all in the plasma-on mode.}
    \label{fig:10}
\end{figure}

The bandwidth–efficiency product is commonly used as an overall figure of merit for electrically small antennas. Figure~\ref{fig:9} presents the simulated product of fractional bandwidth (FBW) and radiation efficiency as a function of $ ka$ for different plasma power levels, alongside the theoretical Chu limit for linear time-invariant ESAs. The results highlight the potential of plasma-matched ESAs to substantially improve the trade-off between antenna size, bandwidth, and efficiency. Note that the plasma sustaining power was not included in the efficiency calculation. In this configuration, antenna efficiency depends on the antenna input power because the plasma requires a minimum sustaining power to achieve a given negative-capacitance state; consequently, the relative impact of plasma power on efficiency decreases as antenna input power increases. Finally, the simulated far-field radiation patterns at 600~MHz are shown in Fig.~\ref{fig:10}. The antenna exhibits the expected doughnut-shaped radiation pattern of a dipole, confirming that plasma integration preserves the fundamental radiation characteristics while enabling wideband impedance matching.

\section{Experimental Results}\label{sec4}

\begin{figure}[!]
    \centering
    \includegraphics[width=0.95\columnwidth]{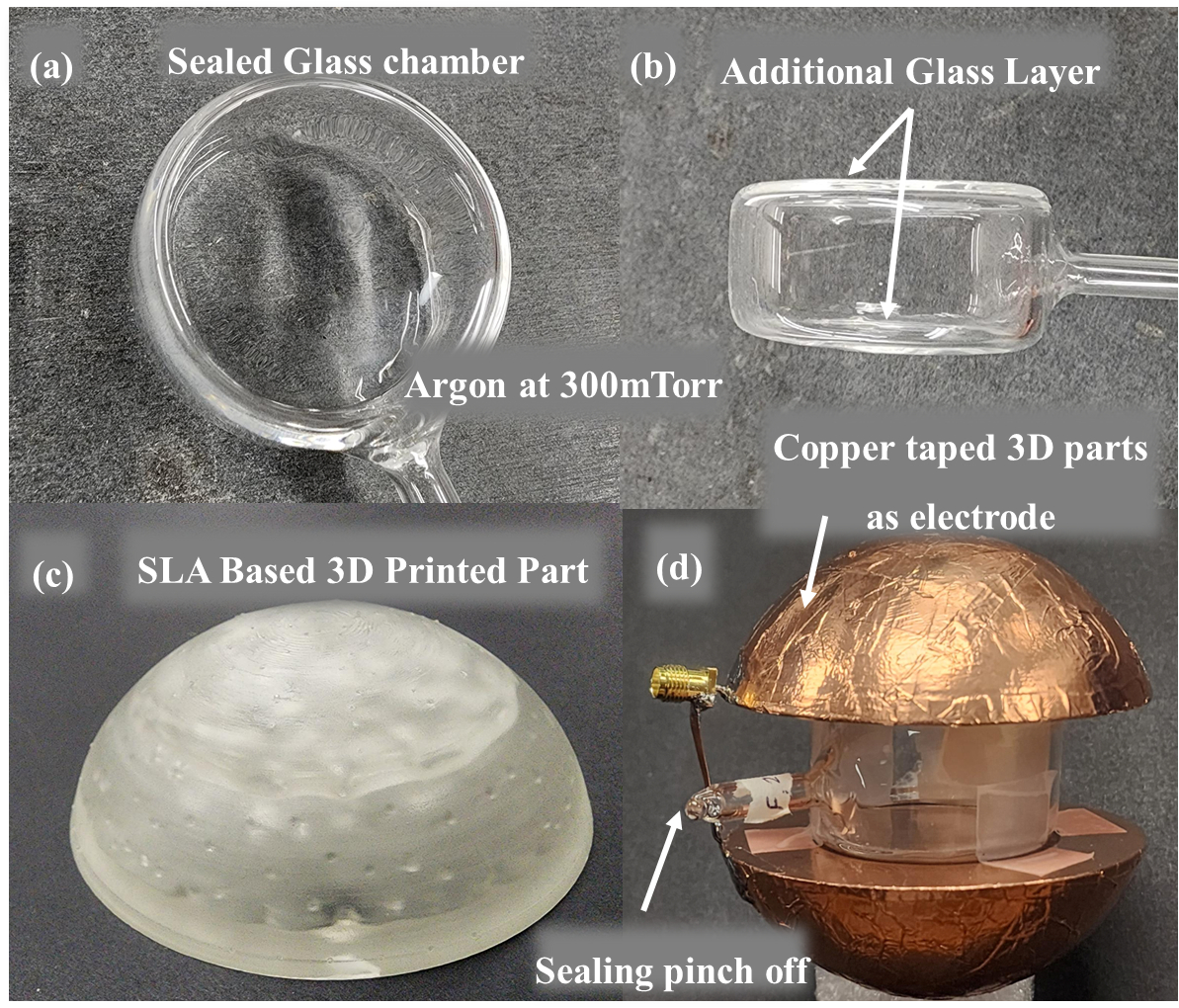}
    \caption{Fabricated components of the plasma-matched ESA: (a) sealed glass plasma cell filled with argon at 300 mTorr, (b) side view of the glass cell showing the additional sealing layer, (c) SLA 3D-printed capped dipole electrodes, and (d) assembled copper-taped capped dipole antenna with the glass plasma cell sandwiched between the electrodes.}
    \label{fig:11}
\end{figure}

The plasma cell was fabricated using borosilicate glass and filled with argon gas at a pressure of 300 mTorr, as shown in Figs.~\ref{fig:11}(a) and \ref{fig:11}(b). The antenna’s domed electrodes were manufactured using 3D stereolithography (SLA) printing, as illustrated in Fig.~\ref{fig:11}(c), and subsequently wrapped with copper tape to form the conductive surfaces. The fully assembled plasma-matched capped dipole antenna is shown in Fig.~\ref{fig:11}(d).

Note that the fabricated prototype deviates slightly from the idealized geometry used in the simulations. This fabrication approach was selected for proof-of-concept validation because of several practical advantages. The structure is straightforward to fabricate, is cost-effective, and avoids complex manufacturing steps such as brazing or CNC machining. The complete assembly weighs approximately 80 g, making it lightweight and easy to handle. However, these advantages come with trade-offs. The use of copper foil results in lower effective electrical conductivity compared to bulk copper, and surface roughness introduced during foil application further degrades conductivity \cite{Rough2,Rough3}. In addition, the presence of a glass layer at the electrode interfaces introduces an additional positive capacitance that must be compensated for by the plasma. This compensation requires a higher plasma excitation power and increases ohmic losses within the discharge.

\begin{figure}[!]
    \centering
    \includegraphics[width=0.98\columnwidth]{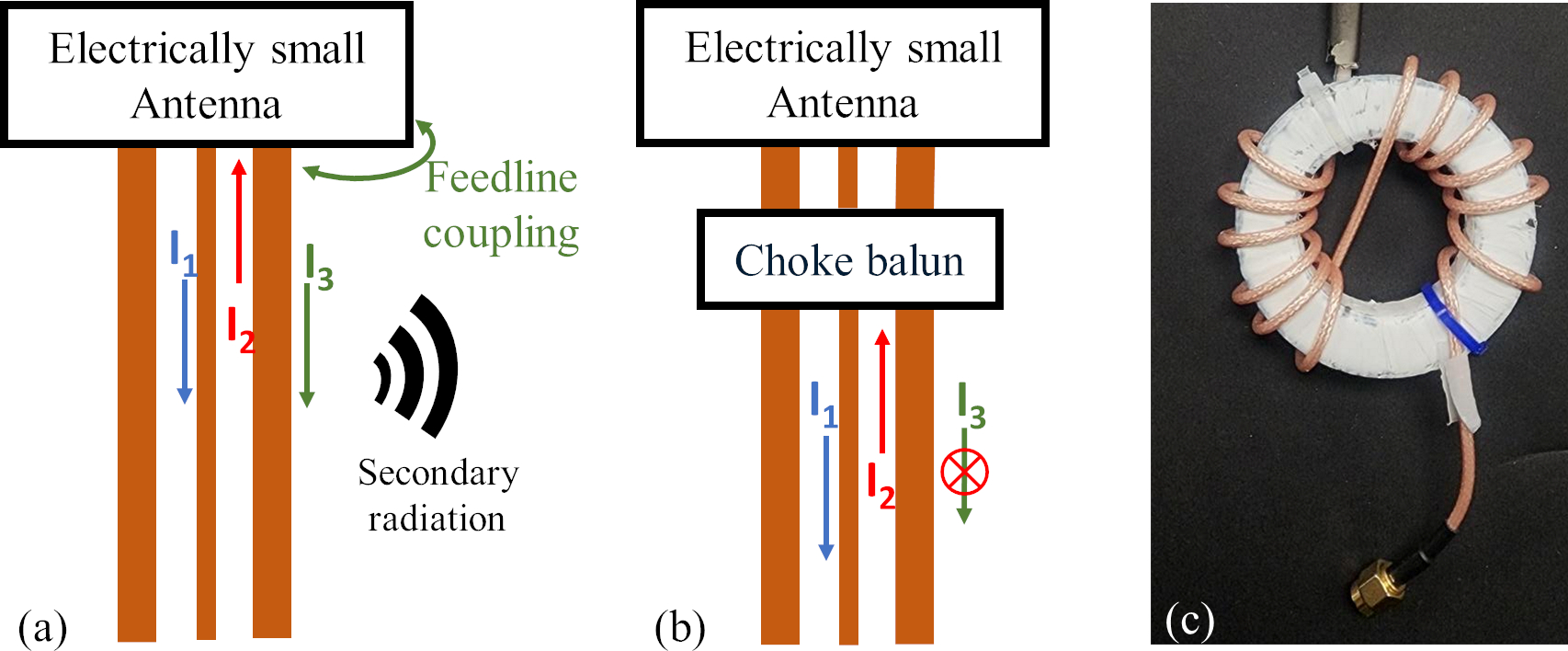}
    \caption{(a) Illustration of the influence of the coaxial feeding cable on measurements of a balanced antenna, (b) application of a choke-based current-balancing balun to suppress secondary radiation, and (c) fabricated toroidal ferrite–based choke balun used in the measurements.}
    \label{fig:12}
\end{figure}

\begin{figure}[!]
    \centering
    \includegraphics[width=0.95\columnwidth]{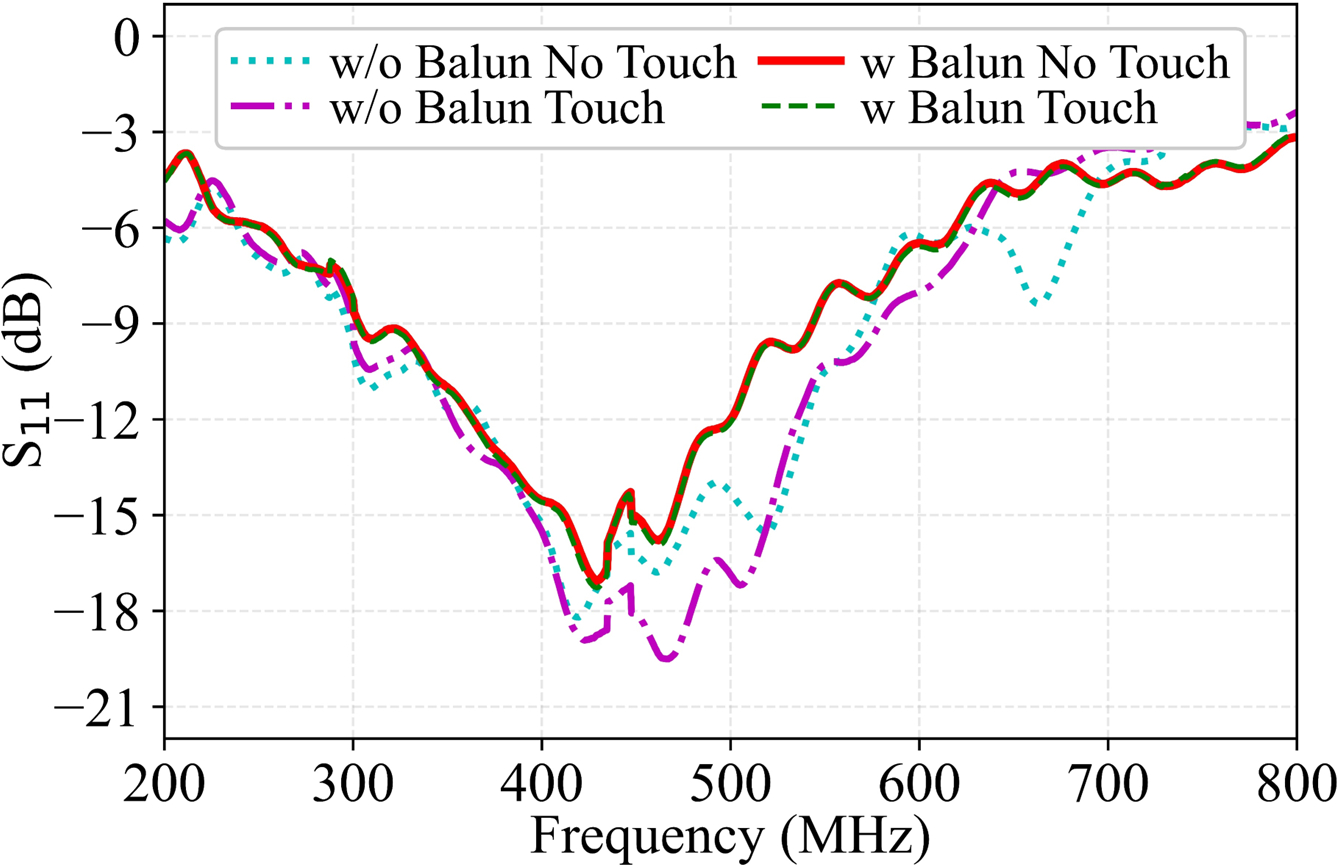}
    \caption{Measured antenna reflection coefficient with and without the current-limiting balun while physically perturbing the feeding cable, demonstrating the effectiveness of the balun in suppressing unwanted common-mode backflow currents.}
    \label{fig:12a}
\end{figure}

Accurate measurement of electrically small antennas, particularly dipole-type ESAs, is inherently challenging due to the influence of the feeding cable. Unwanted common-mode currents can propagate along the outer conductor of the coaxial feed, resulting in secondary radiation that distorts the measured antenna characteristics \cite{SecRed,SecRed1}. These parasitic effects can significantly alter the apparent resonant frequency, input impedance, and radiation pattern, as illustrated schematically in Fig.~\ref{fig:12}(a). To mitigate these effects, a 1:1 choke balun was designed to block the common-mode currents responsible for secondary radiation, as illustrated in Fig.~\ref{fig:12}(b). The fabricated toroidal ferrite-based choke is shown in Fig.~\ref{fig:12}(c). The effectiveness of the choke was experimentally verified by monitoring changes in the measured $S_{11}$ when different sections of the feed cable were physically touched. As shown in Fig.~\ref{fig:12a}, the measured $S_{11}$ remains stable when the choke is present, while significant fluctuations occur in its absence. This confirms the critical role of the choke balun in improving measurement accuracy by suppressing unwanted back-currents. 

\begin{figure}[!]
    \centering
    \includegraphics[width=0.98\columnwidth]{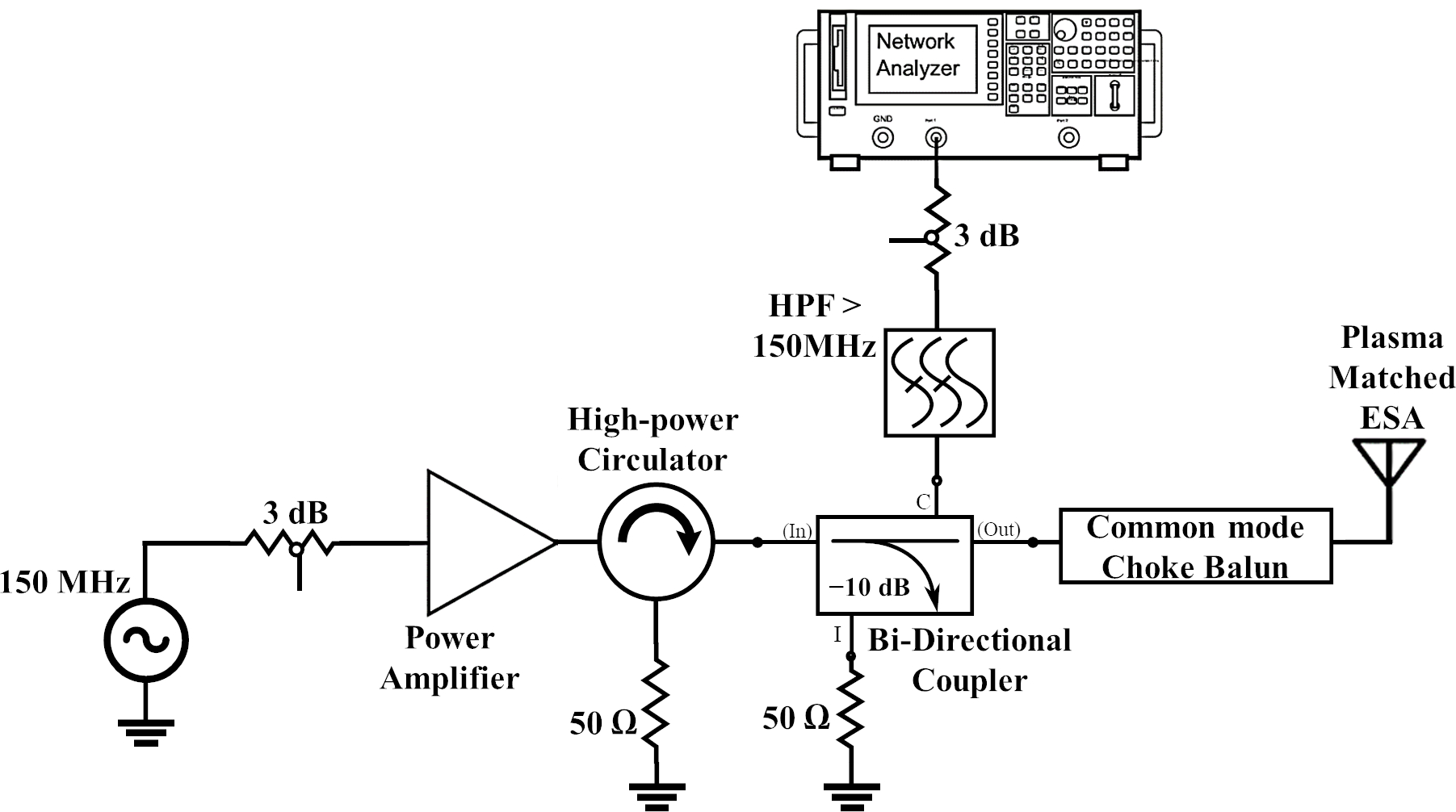}
    \caption{Block diagram of the power delivery and measurement setup, illustrating how the plasma excitation signal and the swept-frequency antenna measurement signal are combined and applied to the single antenna feed port.}
    \label{fig:13}
\end{figure}

The experimental power delivery and measurement setup is shown schematically in Fig.~\ref{fig:13}. The system comprises a directional coupler that enables simultaneous combination of two signals: a low-power swept-frequency signal (200-800 MHz) for antenna characterization and a high-power single-tone signal at 150 MHz for plasma generation. Plasma ignition required approximately 15 W of power, while steady-state operation could be maintained with as little as 2 W. It is important to note that these values represent the electrical power delivered to the system, rather than the net power absorbed by the plasma.

\begin{figure}[!]
    \centering
    \includegraphics[width=0.98\columnwidth]{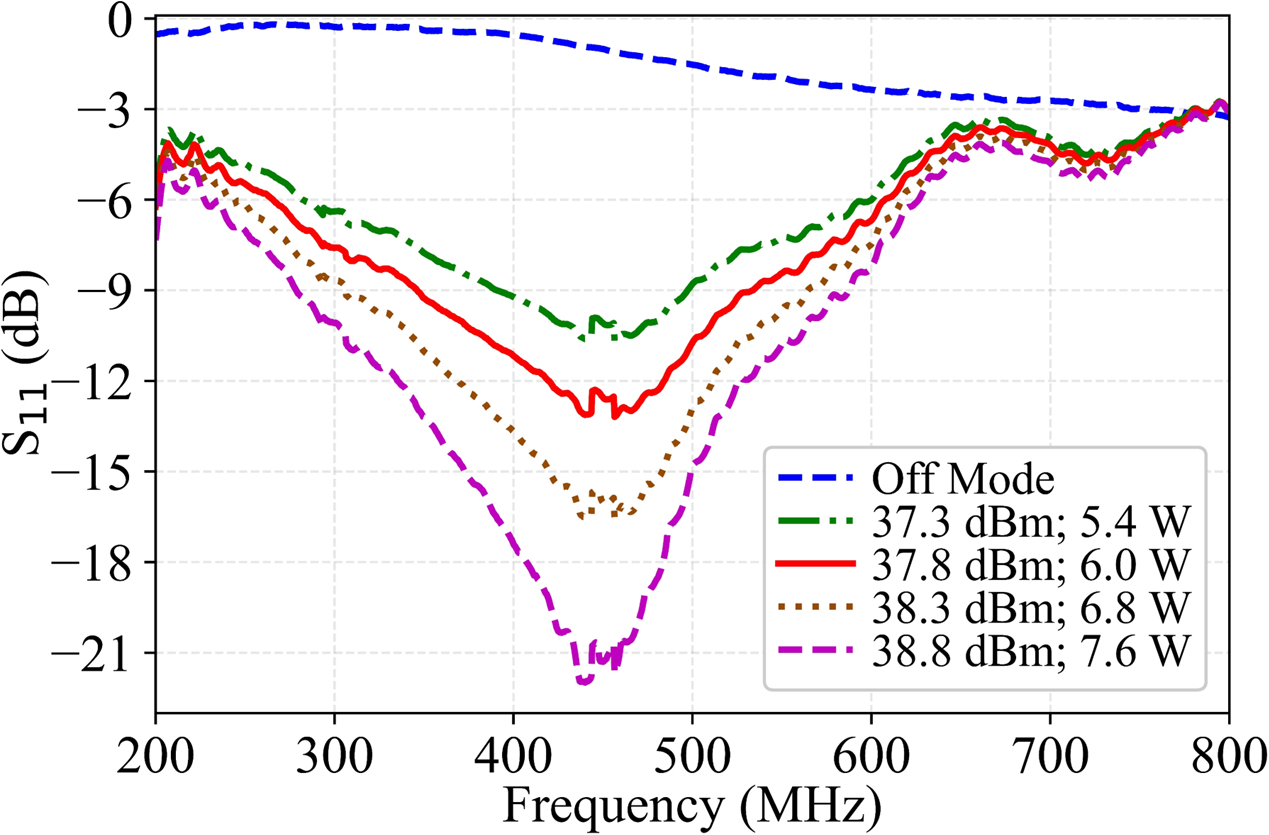}
    \caption{Measured reflection coefficient of the plasma-matched capped dipole antenna at different plasma excitation power levels, demonstrating bandwidth tunability enabled by the plasma; higher plasma density yields broader matching bandwidth at the expense of reduced radiation efficiency.}
    \label{fig:14}
\end{figure}

Figure~\ref{fig:14} presents the measured reflection coefficients of the antenna in both plasma-off and plasma-on states, with plasma input power ranging from approximately 5 to 10 W. To isolate the low-power antenna measurement signal from the high-power plasma excitation signal, a high-pass filter was employed, and $S_{11}$ was measured over the frequency range of 200 to 800 MHz. Compared with the plasma-off condition, the plasma-on state exhibits a substantial improvement in impedance matching, with the reflection coefficient remaining below -6 dB over the frequency range 200-650 MHz. The measured FBW ranges from approximately 72\% to 95\% across plasma input power levels, demonstrating the broadband nature and tunability of the plasma-matching approach.

\begin{figure}[!]
    \centering
    \includegraphics[width=0.95\columnwidth]{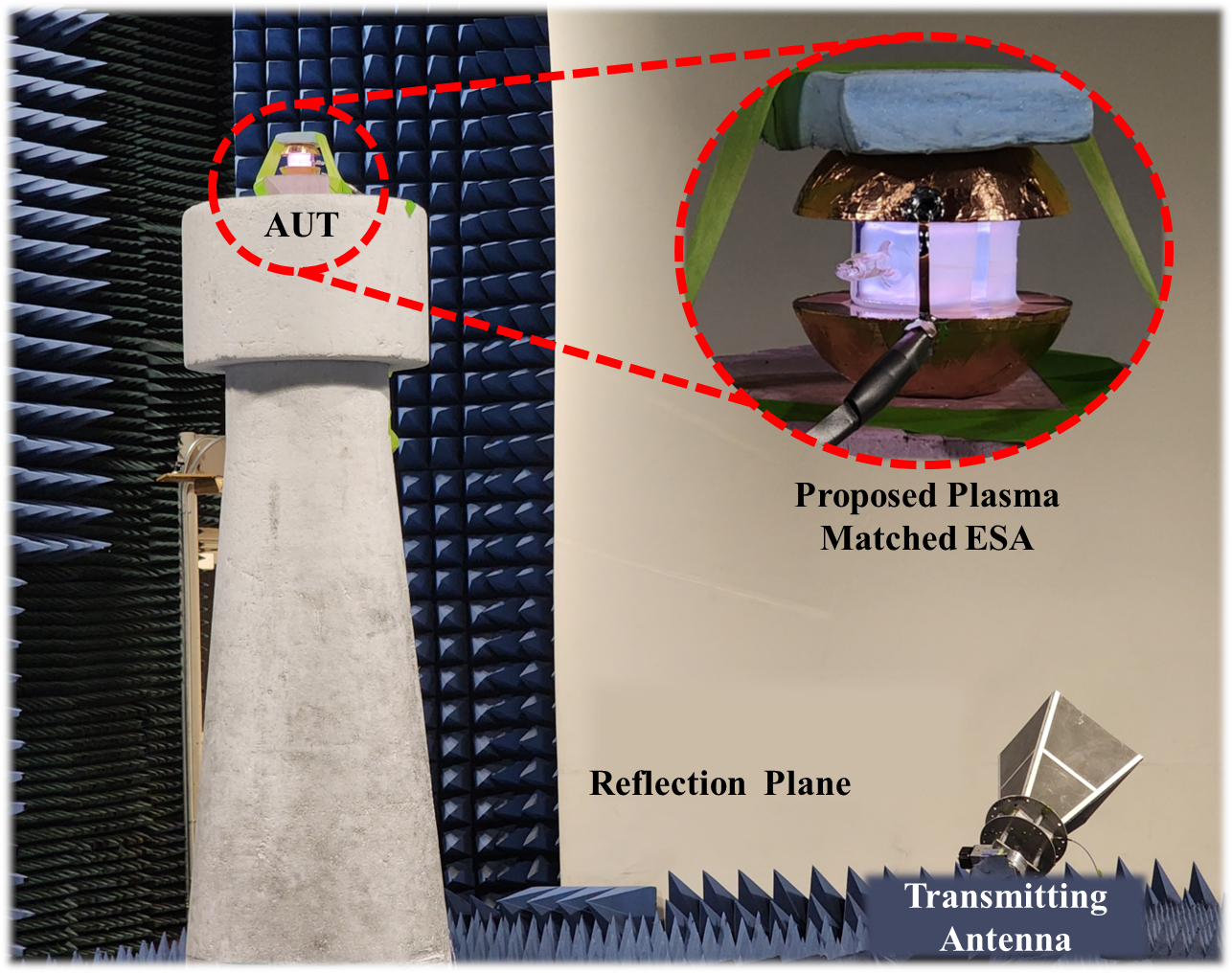}
    \caption{Fabricated plasma-matched electrically small capped dipole antenna under test in the anechoic chamber.}
    \label{fig:15}
\end{figure}

\begin{figure}[!]
    \centering
    \includegraphics[width=0.95\columnwidth]{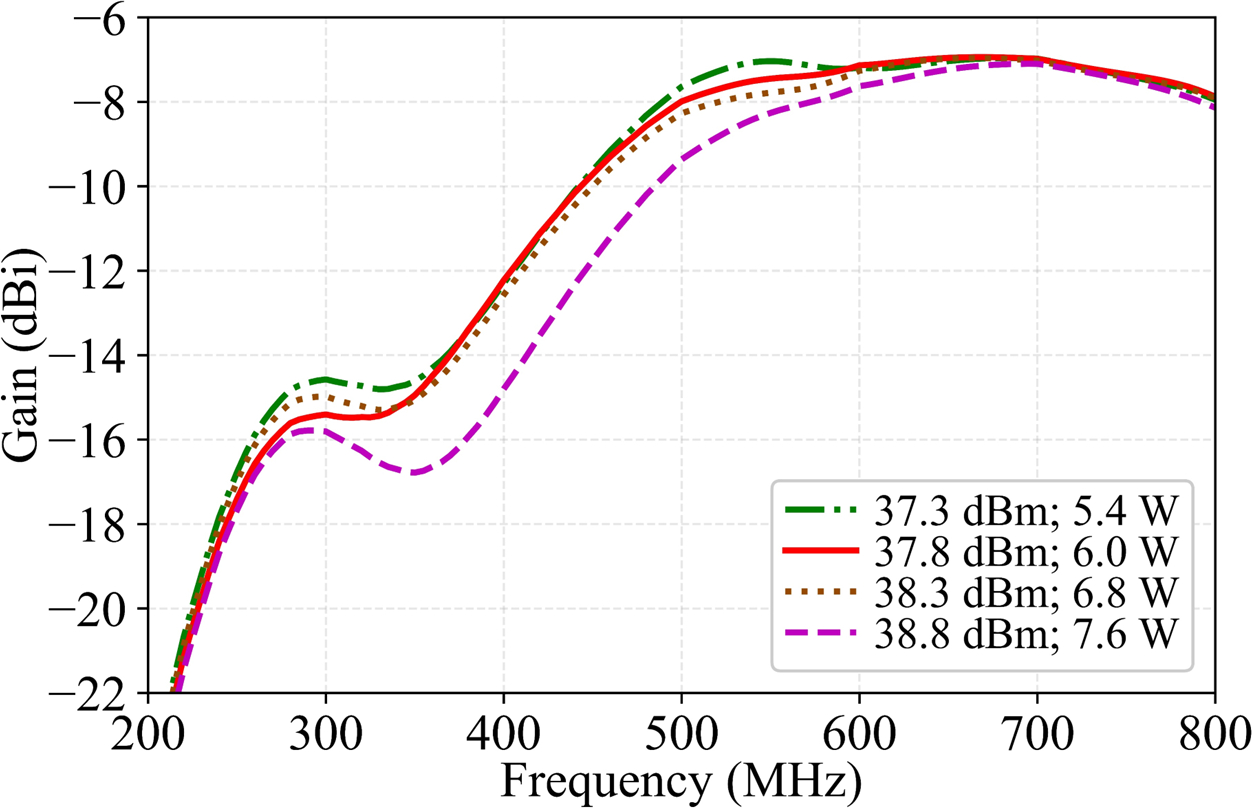}
    \caption{Measured realized gain of the plasma-matched electrically small capped dipole antenna over the 200–800 MHz frequency range for different plasma excitation power levels ranging from approximately 5 to 8 W.}
    \label{fig:16}
\end{figure}

Radiation performance was evaluated in an anechoic chamber, as shown in Fig.~\ref{fig:15}. The measured antenna gain is plotted in Fig.~\ref{fig:16} over the frequency range 200–800 MHz for different plasma power levels. As evident in Figs.~\ref{fig:14} and \ref{fig:16}, a clear trade-off exists between impedance matching bandwidth and radiation efficiency. Increasing the plasma excitation power improves impedance matching by broadening the matched bandwidth; however, this improvement comes with higher ohmic losses due to increased plasma conductivity, reducing the realized gain. These results highlight the need to balance match performance with efficiency when optimizing plasma operating conditions.

\begin{figure}[!]
    \centering
    \includegraphics[width=0.98\columnwidth]{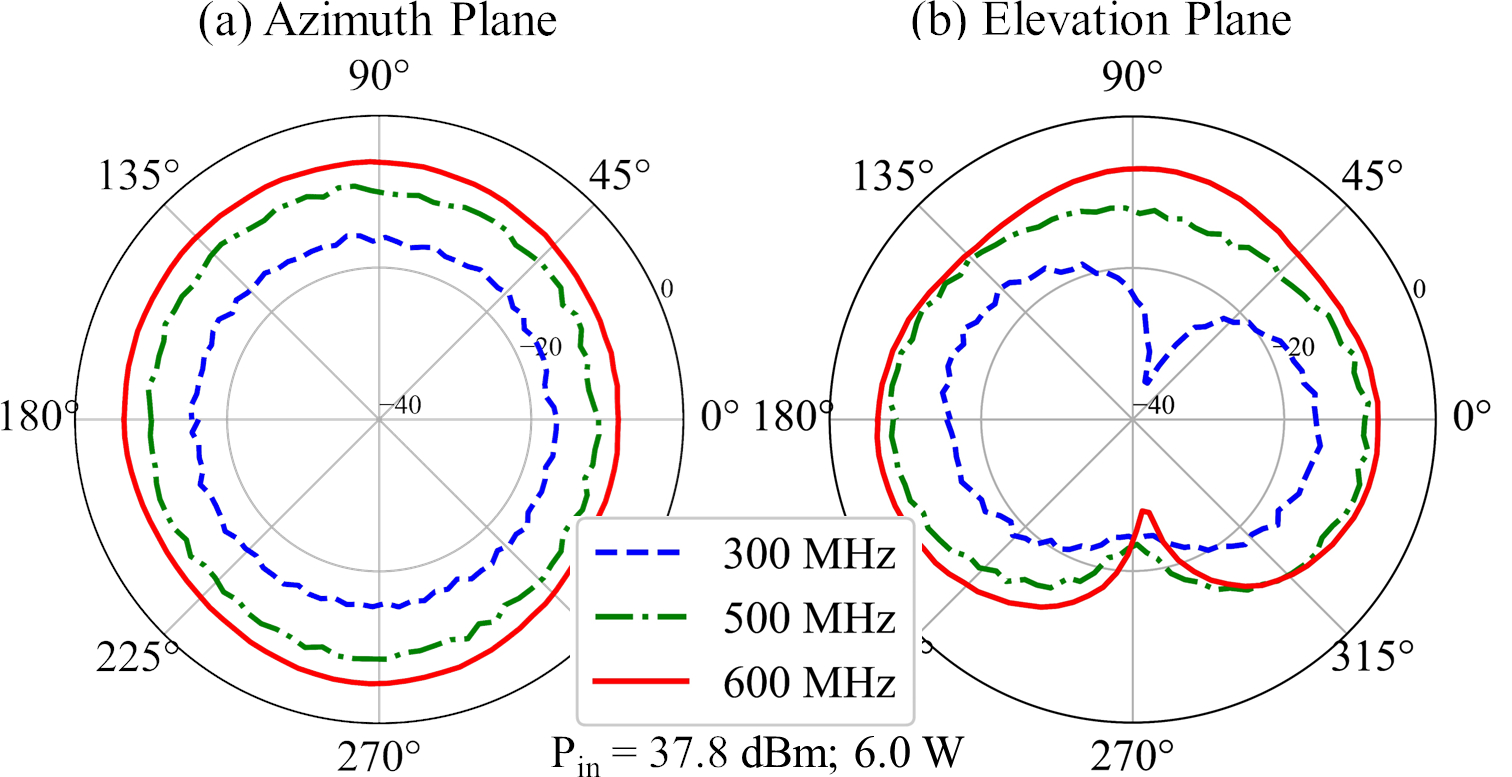}
    \caption{Measured radiation patterns of the plasma-matched capped dipole electrically small antenna in (a) the azimuth plane and (b) the elevation plane at 300, 500, and 600 MHz, shown in blue, green, and red, respectively.}
    \label{fig:17}
\end{figure}

Finally, the measured radiation patterns are shown in Fig.~\ref{fig:17}. In the azimuthal plane, the antenna exhibits a nearly omnidirectional radiation pattern, indicating uniform horizontal coverage. In the elevation plane, the radiation pattern closely resembles that of a conventional dipole antenna, confirming that plasma integration preserves the ESA's fundamental radiation characteristics while enabling wideband impedance matching.

\section{Conclusion}
A plasma-based wideband impedance-matching technique for electrically small antennas has been introduced. By exploiting the inherently negative permittivity of capacitively coupled plasmas to realize a wideband negative capacitor, the proposed approach effectively compensates for the frequency-dependent reactance of ESAs. Using this technique, a prototype ESA achieved better than -6 dB impedance matching across the 150–650 MHz range, corresponding to $ka$ of 0.13 to 0.54. Although experimental performance was limited by practical fabrication constraints, most notably the reduced electrode conductivity and additional positive capacitance introduced by the glass sealing interface, the close agreement between the measured and simulated results validates the robustness and feasibility of the concept. In this technique, plasma excitation power provides a tunable control parameter, enabling dynamic adjustment between bandwidth and radiation efficiency. Although ohmic losses are intrinsic in the mid-collisional plasma regimes, several promising strategies exist to mitigate them, including pulsed plasma excitation and time-gated antenna operation, which will be explored in future work to further enhance radiation efficiency and overall system performance.

\bibliographystyle{IEEEtran}

\bibliography{IEEEabrv,IEEEexample}

@article{TAP1_1,
  title={Fundamental efficiency limits for small metallic antennas},
  author={Pfeiffer, Carl},
  journal={IEEE Transactions on Antennas and Propagation},
  volume={65},
  number={4},
  pages={1642--1650},
  year={2017},
  publisher={IEEE}
}

@article{TAP1_2,
  title={Impedance, bandwidth, and Q of antennas},
  author={Yaghjian, Arthur D and Best, Steven R},
  journal={IEEE Transactions on Antennas and Propagation},
  volume={53},
  number={4},
  pages={1298--1324},
  year={2005},
  publisher={IEEE}
}

@article{TAP1_5,
  title={Non-Foster matched antennas for high-power applications},
  author={Jacob, Minu M and Sievenpiper, Daniel F},
  journal={IEEE Transactions on Antennas and Propagation},
  volume={65},
  number={9},
  pages={4461--4469},
  year={2017},
  publisher={IEEE}
}

@article{TAP1_6,
  title={Wideband, non-Foster impedance matching of electrically small transmitting antennas},
  author={Shih, Ting-Yen and Behdad, Nader},
  journal={IEEE Transactions on Antennas and Propagation},
  volume={66},
  number={11},
  pages={5687--5697},
  year={2018},
  publisher={IEEE}
}

@article{TAP1_9,
  title={On the stability properties of a negative impedance converter},
  author={Brownlie, J},
  journal={IEEE Transactions on Circuit Theory},
  volume={13},
  number={1},
  pages={98--99},
  year={1966},
  publisher={IEEE}
}

@article{TAP1_10,
  title={Non-Foster impedance matching of electrically-small antennas},
  author={Sussman-Fort, Stephen E and Rudish, Ronald M},
  journal={IEEE Transactions on Antennas and Propagation},
  volume={57},
  number={8},
  pages={2230--2241},
  year={2009},
  publisher={IEEE}
}

@article{TAP1_11,
  title={Metamaterial-based efficient electrically small antennas},
  author={Ziolkowski, Richard W and Erentok, Aycan},
  journal={IEEE Transactions on antennas and propagation},
  volume={54},
  number={7},
  pages={2113--2130},
  year={2006},
  publisher={IEEE}
}

@article{TAP1_12,
  title={Application of double negative materials to increase the power radiated by electrically small antennas},
  author={Ziolkowski, Richard W and Kipple, Allison D},
  journal={IEEE Transactions on Antennas and Propagation},
  volume={51},
  number={10},
  pages={2626--2640},
  year={2003},
  publisher={IEEE}
}

@ARTICLE{lit6_1,
  author={Dong, Yuandan and Itoh, Tatsuo},
  journal={Proceedings of the IEEE}, 
  title={Metamaterial-Based Antennas}, 
  year={2012},
  volume={100},
  number={7},
  pages={2271-2285},
  doi={10.1109/JPROC.2012.2187631}}

@inproceedings{lit6,
  title={A planar zeroth-order resonator antenna using a left-handed transmission line},
  author={Sanada, Atsushi and Kimura, Masao and Awai, Ikuo and Caloz, Christophe and Itoh, Tatsuo},
  booktitle={34th European Microwave Conference, 2004.},
  volume={3},
  pages={1341--1344},
  year={2004},
  organization={IEEE}
}

@article{lit7,
  title={Composite right/left-handed transmission line based compact resonant antennas for RF module integration},
  author={Lee, Cheng-Jung and Leong, Kevin MKH and Itoh, Tatsuo},
  journal={IEEE Transactions on Antennas and Propagation},
  volume={54},
  number={8},
  pages={2283--2291},
  year={2006},
  publisher={IEEE}
}

@article{lit10,
  title={Miniaturized substrate integrated waveguide slot antennas based on negative order resonance},
  author={Dong, Yuandan and Itoh, Tatsuo},
  journal={IEEE transactions on antennas and propagation},
  volume={58},
  number={12},
  pages={3856--3864},
  year={2010},
  publisher={IEEE}
}

@article{lit12,
  title={Subwavelength, compact, resonant patch antennas loaded with metamaterials},
  author={Al{\`u}, Andrea and Bilotti, Filiberto and Engheta, Nader and Vegni, Lucio},
  journal={IEEE Transactions on Antennas and Propagation},
  volume={55},
  number={1},
  pages={13--25},
  year={2007},
  publisher={IEEE}
}

@article{lit13,
  title={Electrically small antenna elements using negative permittivity resonators},
  author={Stuart, Howard R and Pidwerbetsky, Alex},
  journal={IEEE transactions on antennas and propagation},
  volume={54},
  number={6},
  pages={1644--1653},
  year={2006},
  publisher={IEEE}
}

@article{lit14,
  title={Resonance and Q performance of ellipsoidal ENG subwavelength radiators},
  author={Ahmadi, Akram and Saadat, Soheil and Mosallaei, Hossein},
  journal={IEEE transactions on antennas and propagation},
  volume={59},
  number={3},
  pages={706--713},
  year={2010},
  publisher={IEEE}
}

@article{lit16,
  title={Narrowband and wideband metamaterial antennas based on degenerate band edge and magnetic photonic crystals},
  author={Volakis, John L and Sertel, Kubilay},
  journal={Proceedings of the IEEE},
  volume={99},
  number={10},
  pages={1732--1745},
  year={2011},
  publisher={IEEE}
}

@article{lit17,
  title={Electrically small split ring resonator antennas},
  author={Alici, Kamil Boratay and Ozbay, Ekmel},
  journal={Journal of applied physics},
  volume={101},
  number={8},
  year={2007},
  publisher={AIP Publishing}
}

@article{lit18,
  title={A folded-monopole model for electrically small NRI-TL metamaterial antennas},
  author={Antoniades, Marco A and Eleftheriades, George V},
  journal={IEEE Antennas and Wireless Propagation Letters},
  volume={7},
  pages={425--428},
  year={2008},
  publisher={IEEE}
}

@article{lit20,
  title={Electrically small, millimeter wave dual band meta-resonator antennas},
  author={Kim, In Kwang and Varadan, Vasundara V},
  journal={IEEE Transactions on Antennas and Propagation},
  volume={58},
  number={11},
  pages={3458--3463},
  year={2010},
  publisher={IEEE}
}

@article{lit22,
  title={Planar artificial magnetic conductors and patch antennas},
  author={Zhang, Ying and Von Hagen, J{\"u}rgen and Younis, Marwan and Fischer, Christian and Wiesbeck, Werner},
  journal={IEEE Transactions on Antennas and Propagation},
  volume={51},
  number={10},
  pages={2704--2712},
  year={2003},
  publisher={IEEE}
}

@article{lit23,
  title={Reflection phase characterizations of the EBG ground plane for low profile wire antenna applications},
  author={Yang, Fan and Rahmat-Samii, Yahya},
  journal={IEEE Transactions on antennas and propagation},
  volume={51},
  number={10},
  pages={2691--2703},
  year={2003},
  publisher={IEEE}
}

@article{lit24,
  title={Artificial magnetic conductor surfaces and their application to low-profile high-gain planar antennas},
  author={Feresidis, Alexandros P and Goussetis, George and Wang, Shenhong and Vardaxoglou, John C},
  journal={IEEE Transactions on Antennas and Propagation},
  volume={53},
  number={1},
  pages={209--215},
  year={2005},
  publisher={IEEE}
}

@article{lit28,
  title={High efficiency bandwidth VHF electrically small antennas through direct antenna modulation},
  author={Santos, Jean Paul Dytioco and Fereidoony, Foad and Hedayati, Maziar and Wang, Yuanxun Ethan},
  journal={IEEE Transactions on Microwave Theory and Techniques},
  volume={68},
  number={12},
  pages={5029--5041},
  year={2020},
  publisher={IEEE}
}

@article{lit29,
  title={Space-Time Modulation of a Multimode Electrically Small Antenna for Increased Matching and Efficiency Bandwidths},
  author={Fritts, Zachary and Babaee, Amirhossein and Young, Steve M and Grbic, Anthony},
  journal={IEEE Transactions on Antennas and Propagation},
  year={2024},
  publisher={IEEE}
}

@article{TAP1_13,
  title={A comprehensive circuit modeling approach for self-sustained capacitively coupled microwave plasmas},
  author={Ramesh, Sandeep Narasapura and Semnani, Abbas},
  journal={IEEE Transactions on Plasma Science},
  volume={49},
  number={9},
  pages={2690--2699},
  year={2021},
  publisher={IEEE}
}

@inproceedings{TAP1_15,
  title={A Plasma-Based Technique for Wideband Matching of Electrically Small Antennas},
  author={Semnani, Abbas and Singhal, Kushagra and Moon, Samsud T},
  booktitle={2023 IEEE International Symposium on Antennas and Propagation and USNC-URSI Radio Science Meeting (USNC-URSI)},
  pages={1349--1350},
  year={2023},
  organization={IEEE}
}

@article{SecRed,
  title={Low-profile, multi-element, miniaturized monopole antenna},
  author={Hong, Wonbin and Sarabandi, Kamal},
  journal={IEEE Transactions on Antennas and Propagation},
  volume={57},
  number={1},
  pages={72--80},
  year={2009},
  publisher={IEEE}
}

@ARTICLE{SecRed1,
  author={Rao, Menglou and Sarabandi, Kamal},
  journal={IEEE Transactions on Antennas and Propagation}, 
  title={A Tunable, High-Gain, Very Low-Profile Composite Monopole Antenna for Low-Frequency Applications}, 
  year={2018},
  volume={66},
  number={7},
  pages={3286-3294},
  doi={10.1109/TAP.2018.2826653}}

@ARTICLE{comsol,
    year={COMSOL Multiphysics® v. 6.4. www.comsol.com. COMSOL AB, Stockholm, Sweden.}}

@ARTICLE{Rough2,
  author={Holloway, C.L. and Kuester, E.F.},
  journal={IEEE Transactions on Microwave Theory and Techniques}, 
  title={Power loss associated with conducting and superconducting rough interfaces}, 
  year={2000},
  volume={48},
  number={10},
  pages={1601-1610},
  }

@ARTICLE{Rough3,
  author={Pucel, R.A. and Masse, D.J. and Hartwig, C.P.},
  journal={IEEE Transactions on Microwave Theory and Techniques}, 
  title={Losses in Microstrip}, 
  year={1968},
  volume={16},
  number={6},
  pages={342-350},
  doi={10.1109/TMTT.1968.1126691}}

@book{PlasmaDischarge,
  title={An introduction to gas discharges: Pergamon international library of science, technology, engineering and social studies},
  author={Howatson, Alastair Macrae},
  year={2013},
  publisher={Elsevier}
}

@article{PlasmaChem3,
  title={Simulation of argon-excited microwave plasma reactor for green energy and CO2 conversion application},
  author={Ong, Mei Yin and Chia, Shir Reen and Milano, Jassinnee and Nomanbhay, Saifuddin and Chew, Kit Wayne and Yusaf, Talal and Show, Pau Loke},
  journal={Applied Energy},
  volume={353},
  pages={122160},
  year={2024},
  publisher={Elsevier}
}

@article{PlasmaChem6,
  title={Analysis of hollow cathode aperture wall profile on pseudospark discharge-based sheet electron beam source},
  author={Singhal, Kushagra and Gurjar, Nikita and Jain, Sahil and Abhishek, Anand and Starodubov, AV and Ryskin, NM and Kumar, Niraj and others},
  journal={IEEE Transactions on Plasma Science},
  volume={49},
  number={1},
  pages={302--306},
  year={2020},
  publisher={IEEE}
}

@ARTICLE{plasESA1,
  author={Chen, K.M. and Lin, C.C.},
  journal={Proceedings of the IEEE}, 
  title={Enhanced radiation from a plasma-imbedded antenna}, 
  year={1968},
  volume={56},
  number={9},
  pages={1595-1597},
  doi={10.1109/PROC.1968.6659}}

@article{plasESA2,
  title={Towards antenna miniaturization at radio frequencies using plasma discharges},
  author={Laquerbe, Vincent and Pascaud, Romain and Laffont, Adrien and Callegari, Thierry and Liard, Laurent and Pascal, Olivier},
  journal={Physics of Plasmas},
  volume={26},
  number={3},
  year={2019},
  publisher={AIP Publishing}
}

\end{document}